*REVIEW*

# Review on the astronomical estimators of number of flare stars

A. A. Akopian

*NAS RA V. Ambartsumian Byurakan Astrophysical Observatory (BAO)*
e-mail: *aakopian@bao.sci.am*

## Abstract

Review is devoted to estimators introduced in astronomy by Ambartsumian and his followers that are used to estimating the unknown number of flare stars and other randomly flashing objects (stars with X-ray flares, solar type stars with superflares). Some important astronomical applications of them are presented. The development of these methods in astronomy have proceeded regardless of development of analogous methods of mathematical statistics. Brief history of this development and parallels with similar statistical methods is presented.

## 1. Introduction

In 1968, at the conference devoted to his 60 anniversary, famous soviet and Armenian astronomer Ambartsumian (1969) has given an unforeseen report in which has offered subtle and unexpectedly simple method for estimating the total number of flare stars in star systems. This review is devoted to 50 years jubilee of Ambartsumian's estimator, which is in widely usage now in various branches of statistics, medical and social sciences, biology and ecology, linguistics, criminology, epidemiology, cryptography, etc. In these branches this method, or closely related methods, named by different ways: capture-recapture, mark recapture, capture-mark-recapture, mark-recapture, sight-resight, mark-release-recapture etc.

Capture-recapture method is widely used to estimate the unknown size of a population. In practice, usually it is impossible to count directly every individual member in a population because of time limit or other constraints. Therefore, in many situations, capture-recapture and similar methods produce a statistically valid estimate of a population size in a more efficient and timely manner than a direct counting. The idea of the capture-recapture technique can be traced back to a 1786 paper by Laplace (Seber, 1982) who applied the method in1802 to estimate the population of France (Stigler, 1986), but only in last decades these methods systematically have been developed and used.

There are some excellent books and reviews on this topic (e.g., Amstrup et al. (2005),McCrea and Morgan (2014), Bohning et al. (2017) ). But in all these books and reviews was missed the fact that such techniques have been in use in astronomy over 50 years and that one of the most important and frequently used estimator of the unknown size of a population(known as Chao estimator) firstly was suggested and applied in astronomy.

Since 1968 Ambartsumian estimator many times was used to estimate the unknown number of flare stars and other randomly flaring objects, such as the stars with X-ray flares, solar type stars with power flares (superflares). Perhaps, with the accumulation of necessary data, it can be successfully used in studies of supernova stars and gamma bursts.

Results obtained by Ambartsumian stimulated works in the study and statistics of flare stars. In particular, some other new estimators of total number of flare stars were suggested. It is important to note that the development and application of these methods in astronomy took place in parallel but independently of similar methods developed in mathematical statistics. This led to a situation that some of mathematical results have been obtained in both fields of science, sometimes with significant difference over time.

The objective of this review is to present estimators suggested by astronomers with some applications in astronomy, brief history of the development of these methods in astronomy and parallels with similar statistical methods.

## 2. Ambartsumian estimator. First estimation of the total number of flare stars.

Flare stars are stars that show sudden, short-duration flares with amplitudes ranging from a few tenths of magnitude to $6^m-7^m$ magnitudes at visible wavelengths. Flare stars are red dwarf stars, which have a cool photosphere at 2500 – 4000K. The flare rise time is a few seconds to a few tens of seconds, and the decline lasts a few minutes to a few tens of minutes. At maximum optical brightness the larger flares may contribute more than 100times the total output of the quiescent star. The flare mechanism is not well understood yet. These stars are closely associated with nebulae and are found in the regions of the Orion association, the Pleiades and other galactic open clusters.

Up to now the only way to discovering and identify them as flare stars is to register a flare, and a question arises, how many undiscovered flare stars exist in these systems? The key for solving this problem was given in 1968 by Ambartsumian. An estimation o ftotal number of flare stars in any stellar system has been obtained (Ambartsumian, 1969) under two following assumptions:

1. The sequence of flares at each flare star represents a random poissonian process.
2. All flare stars of the given system have same mean frequency of flares.

In frames of adopted approach, the number of stars $n_k$ at which showed $k$ flares is defined by expression:

$$n_k = Np_k, \quad p_k = \frac{(\nu t)^k e^{-\nu t}}{k!} \quad (1)$$

where N –the full number of flare stars in a system, ν - mean frequency of flares, t - the effective time duration of observation of the system. The relation (1) allows to estimate number $n_0$ of flare stars flares of which have not been registered yet, through the known numbers $n_1$ and $n_2$ of stars which already showed, accordingly, one and two flares, as follows:

$$n_0 = \frac{n_1^2}{2n_2} \quad (2)$$

It should be noted that the mentioned above first assumption is quite well-founded. The possibility of expressing the sequence of flares at individual stars through Poisson law has been confirmed in (Oskanyan and Terebizh, 1971), based on the study of the longtime series of photoelectric observations of some flare stars in the vicinity of Sun.

As to the assumption 2, one can give it up. Chavushian (1979) checking the behavior of the following observed sequences

$$\frac{2n_2}{n_1}, \frac{3n_3}{n_2}, \ldots \frac{kn_k}{n_{k-1}} \quad (3)$$

confirmed the heterogeneity of Pleiades cluster flare stars flares frequency. Ambartsumianet al. (1970) using by Cauchy-Schwarz inequality have shown that when assumption 2 is incorrect, the relation (2) turns into an inequality and its application gives the lower bound of the number $n_0$. The upper bound of number $n_0$ has been estimated (Ambartsumian et al.,1970) in the assumption that the density of distribution of frequencies of flares is too slowly decaying exponential function. As a result, takes place inequalities:

$$\frac{n_1^2}{n_2} \geq n_0 \geq \frac{n_1^2}{2n_2}. \tag{4}$$

The (2) takes place only between mathematical expectations of $n_0$, $n_1$ and $n_2$. For want of something better, for estimation of n0 in (2) instead of means $n_1$ and $n_2$ their observed values are used.

At first time, this estimator was applied to estimate the total number of flare stars in Pleiades cluster, which was most well studied flare stars system at that time. Having estimate of the total number of stars the Pleiades cluster, Ambartsumian came to an important cosmogonical conclusions concerning to these stars. The first of those, related to the abundance of flare stars in the Pleiades cluster, is a decisive argument supporting the fact that the stage of flare activity is a regular stage in the evolution through which all the red dwarf stars must pass. Later, these results were confirmed by further research and observations of flare stars, which were held in observatories of Byurakan (Armenia), Tonantzintla (Mexico),Asiago (Italy), etc., within the framework of international scientific cooperation. Morethan 500 flare stars have been revealed in the Orion association and approximately the same number in the Pleiades cluster. Estimated total number in the observed region of Orion association is 1400÷1500, in Pleiades cluster - 900÷1000 stars. Flare stars have also been discovered in other stellar clusters (Praesepe, Hyades, etc.). Numerous earlier applications of Ambartsumian estimator concerning to flare stars, were summarized in some books and reviews, e.g., Mirzoyan (1981).

## *2.1. Recent applications of Ambartsumian estimator*

Recently it was used in studies of new flare stars system in open cluster M37 (Chang et al., 2015) and other celestial objects that are closely related to the flare stars and flare phenomenon (Akopian, 2013, 2017; Pye et al., 2015).

Based on one-month long MMT time-series observations of the open cluster M37, Chang et al. (2015) monitored light variations of nearly 2500 red dwarfs and successfully identified420 flare events shown by 312 cluster M dwarf stars. Using the Ambartsumian estimator, two cases have been considered (both homogeneity-case I and heterogeneity-case II of flare frequency). For these two cases, Chang et al. (2015) estimated the total number of flare stars N and obtained the two parameters $n_0$, N for both case I ($n_0$ = 442, N = 749) and case II ($n_0$ = 510, N = 817), respectively. For members of M37 cluster with high membership probability ($p_{mem}>0.2$) for case I $n_0=$ 114, N = 194 and for case II n= 166, N = 246, respectively.

Pye et al. (2015), using by XMM-Newton Serendipitous Source Catalogue and its associated data products, present a uniform, large-scale survey of X-ray flare emission. These rendipitous sample demonstrates the need for careful calculation of flaring rates, especially when normalizing the number of flares to a total exposure time, where it is important to consider both the flared stars and those from which flares was not detected, since in survey, the latter outnumber the former by more than a factor ten. Pye et al. (2015) emphasize that they are not implying that the stars, which have yet not produced detectable flares in the 2XMM survey observations, are intrinsically without flare activity, and note that this result is broadly consistent with estimates based on Poisson statistics.

More recently, the Ambartsumian estimator was applied to solar-type stars with very powerful flares (called superflares). A superflare on a solar type star is a brightening that has an energy of from $10^{33}$ to more than $10^{38}$ erg and lasts from a few minutes to several days(Schaefer, 2012). In comparison, the largest flare ever observed on the Sun - the 1859 Carrington event - had a total energy of about $10^{32}$erg. The increasing interest in superflares is related with the rapid development of astrobiology and research on exoplanets. On one hand, the huge output energy of superflares may make any planet around these stars unsuitable for life. On the other hand, superflares may serve as the source of high energy radiation needed for the formation of organic molecules, and it is possible that planetary systems surrounding "superflare" stars are a good place to search for exoplanetary life that has escaped the destructive consequences of huge flares in its subsequent evolution (Schaefer,2012).

Another key issue is capable our Sun to produce such superflares, which can be dangerous for the existing forms of life on Earth? There are two opposite opinion about this possibility. Without going into astrophysical details, we note that in order to answer to this and other questions related to the nature superflares, one should determine at first the approximate size of the sample of stars capable to produce superflares.

Considering the data obtained by the orbital observatory "Kepler", Maehara et al.(2012) have registered 365 superflares on 148 stars, including some from slowly rotating solar-type stars, from about 83,000 stars observed over 120 days. With these data it was estimated the number of stars producing superflares (Akopian, 2013). Later Shibayama et al.(2013) studies the same sample of stars with an observation period of roughly four times longer (about 500 days). Altogether there were found 279 stars with the 1547 superflares.

These data provide more accurate estimation of total number of superflare producing stars (Akopian, 2017). Fig.1 shows the temporal behavior of estimate of the total number N of stars producing superflares obtained by Ambartsumian estimator (both lower and upper bound). For comparison same estimate obtained by Zelterman (1988) estimator (see below,possible upper bound (Bohning, 2010)) also given in fig.1. It is easy to see that only ~0.5%of solar-type stars are able to produce superflares, that probably means that they have some yet unknown rare features or they are in very short stage of evolution. Clarification of these features will help to answer the question about possibility of superflares on the Sun.

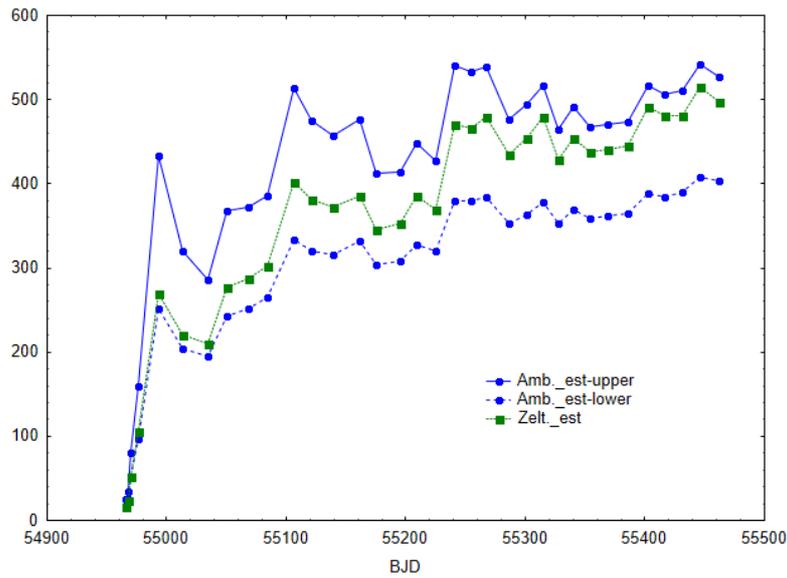

**Figure 1**. The temporal behavior of estimate of the total number of solar type "superflare" stars. On the X-axes barycentric julian days (BJD) is given.

## *2.2 Some mathematical properties of Ambartsumian estimator*

• When assumption 2 (see above) is incorrect, Ambartsumian estimator (2) gives the lower bound of the number $n_0$ (Ambartsumian et al., 1970).

• Ambartsumian estimator is robust one. This important result was obtained by R. Ambartsumian (1998) who has shown, that among possible estimators of type

$$n_0(t) = c_{k,l,m}(t) \frac{n_k(t) n_l(t)}{n_m(t)},  \quad (5)$$

where the factor $c_{k,l,m}$ generally depends on kind of density distribution of flares frequency, robust estimators are those for which equality k + l = m takes place; i.e. the estimator (2) is robust. Hence, according to R. Ambartsumian (1998) another robust estimator is

$$n_0(t) \sim \frac{n_1(t)n_2(t)}{n_3(t)}. \tag{6}$$

- Variation of total number $N = n_1^2/2n_2 + N_1$ were derived by Chao and improved by Bohning (2010) is equal:

$$\sigma_N^2 = n_2 \left[ \frac{1}{2}\left(\frac{n_1}{n_2}\right)^2 + \left(\frac{n_1}{n_2}\right)^3 + \frac{1}{4}\left(\frac{n_1}{n_2}\right)^4 - \frac{1}{4}\frac{n_1^4}{N_1 n_2^3} - \frac{1}{2}\frac{n_1^4}{n_2^2(2n_2 N_1 + n_1^2)} \right]$$

Variation of $n_0 = n_1^2$ has following form:

$$\sigma_{n_0}^2 = \frac{n_1^3}{n_2^2}\left(1 + \frac{1}{4}\frac{n_1}{n_2}\left(1 - \frac{n_2}{N_1}\right)\right)$$

- The mean flare frequency can be determined from the expression (Ambartsumian, 1969):

$$\langle \nu t \rangle = \frac{2n_2}{n_1}. \tag{7}$$

which is a maximum likelihood estimate for the doubly truncated Poisson distribution in which no events with k = 0 and k ≥3 are observed (Akopian, 1998b,a).

## 3. Other estimators of total number of flare stars

Results obtained by Ambartsumian inspired works in the study and statistics of flare stars. In particular, some other new estimators of total number of flare stars were suggested. Most of them rarely used during this time, as it turned out that of Ambartsumian estimator is more simple and mathematically well founded and better suited for such kind of task. In the paper by Mirzoyan et al. (1977) the following estimator has been suggested:

$$N = \frac{\sum_{k=1} n_k}{1 - e^{-\langle \nu t \rangle}} = \frac{N_1}{1 - e^{-\langle \nu t \rangle}}, \tag{8}$$

where $\langle \nu t \rangle$ is defined as the solution of transcendental equation:

$$\frac{\nu t}{1 - e^{-\nu t}} = \frac{\sum_{k=1} k n_k}{\sum_{k=1} n_k} = \frac{n}{N_1}. \tag{9}$$

Here after $n \equiv \sum_{k=1} k n_k$ - total number of flares, $N_1 \equiv \sum_{k=1} n_k$ - total number of stars with one or more flares (discovered flare stars).

Later the same estimator has been suggested by Akopian (1998b), where other estimatorson the basis of poissonian distribution have been suggested also. Well known estimatorsof parameter vt of truncated at point k = a poissonian distribution have been used. Theestimators of number of flare stars have been obtained from the assumption of equality ofthese estimators and the estimators of non-truncated poissonian distribution:

$$\langle vt \rangle_{tr} = \langle vt \rangle_{ntr} \equiv \frac{n}{N}. \tag{10}$$

Truncated at point k = a Plackett unbiased estimator $\langle vt \rangle_{tr} = \frac{\sum_{k=a+2} k n_k}{\sum_{k=a+1} n_k}$ (Plackett, 1953),truncated at point k = 0 maximum likelihood biased estimator $\frac{\langle vt \rangle_{tr}}{1-exp(-\langle vt \rangle_{tr})} = \frac{\sum_{k=1} k n_k}{\sum_{k=1} n_k}$ (David and Johnson, 1952) and truncated at point k = 0 unbiased estimator with minimumvariance $\langle vt \rangle_{tr} = C \frac{\sum_{k=1} k n_k}{\sum_{k=1} n_k}$ (Tate and Goen, 1958), where C expressed by means of Stirlings numbers, have been used.

Estimator (8) for the unknown number of stars can be obtained from maximum likelihood estimator. From Plackett unbiased estimator the following estimator is obtained:

$$n_0 = n_1 \frac{\sum_{k=a+1} n_k}{\sum_{k=a+2} k n_k}. \tag{11}$$

When $a = 0$ (most important and applicable case):

$$n_0 = n_1 \frac{N_1}{n - n_1} = n_1 \frac{(n_1 + n_2 + \cdots + n_k + \ldots)}{2n_2 + 3n_3 + \ldots k n_k + \ldots} \tag{12}$$

$$N = N_1 \frac{n}{n - n_1}. \tag{13}$$

When the flares mean frequencies of flare stars are different, the estimator (11) also gives the lower bound for number of unknown flare stars (Akopian, 1998b,a). Because of complete proofing given in (Akopian, 1998a) is not accessible, we give it in appendix A. Interesting and simple estimator of total number star is obtained from minimal variance($a = 0$) estimator of vt:

$$N = \frac{S_{St}(n, N_1)}{S_{St}(n-1, N_1)} \tag{14}$$

where $S_{St}(x, y)$ is the Stirling number of second kind (in this task - number of ways to partition a set of x flares into y discovered flare stars).

***Prediction of number of new flare stars.*** In the statistics of star flares the prediction problem has been considered by Mnatsakanian (1986) and Mnatsakanian and Mirzoyan(1988). According to Mnatsakanian and Mirzoyan (1988), if $n_k(T)$ are the known numbers of flare stars with k flares at the moment T, then the same numbers at the moment t can be estimated by following way:

$$n_r(t) = \sum_{k=0}^{\infty} n_k(T) \cdot C_k^r \left(\frac{t}{T}\right)^r \left(1 - \frac{t}{T}\right)^{k-r}, \quad r = 0, 1 \ldots \tag{15}$$

In particular,

$$n_0(t) = \sum_{k=0}^{\infty} n_k(T) \cdot \left(1 - \frac{t}{T}\right)^k = \sum_{k=0}^{\infty} n_k(T)(-1)^k \cdot \left(\frac{t-T}{T}\right)^k \tag{16}$$

## 4. Comparison and parallels with analogous statistical estimators

As mentioned above, in last decades statistical methods of estimation of unseen objects rapidly developed and now there are the set of estimator analogous of astronomical. Some of them are similar or tightly linked with astronomical estimators. However, it must be noted that in general case the reasons and the ways of obtaining these identical results were different.

Lets note some of them:

• Now Ambartsumian estimator (Ambartsumian, 1969; Ambartsumian et al., 1970) is well known among statisticians as Chao estimator. 15 years after Ambartsumian's works Chao (1984) suggested same estimator and showed (17 years after Ambartsumian et al. (1970)) that this estimator gives of lower bound (Chao, 1987, 1989). In later works, Chao and his colleagues proposed some modified and adapted to different tasks versions of the estimator, established of some their statistical properties and applied them.

• It seems, that among the other estimators, mentioned higher Zelterman estimator (Zelterman, 1988) can be useful for astronomical purposes, as an only alternative to Ambartsumian upper bound estimate, since according to Bohning (2010) under some realistic conditions it gives upper bound estimate. The conventional Zelterman estimator can be presented in following form:

$$N = \frac{N_1}{1 - exp(-\langle \nu t \rangle)},$$

where $\langle \nu t \rangle = 2n_2/n_1$. In generalized form of Zelterman estimator other estimate of $\langle \nu t \rangle$ (e.g. $\sum_{k=1}^{l}(k+1)n_{k+1}/\sum n_k$) can be used. It is easy to see that Zelterman estimator almost identical of estimator (8) earlier proposed by Mirzoyan et al. (1977).

• Estimator (11) (Akopian, 1998b) in the particular case $a = 0$ (see (13)) coincides with the Good-Turing estimator (Good, 1953), referred also as Darroch-Ratcliff estimator (Darroch and Ratcliff, 1980). This estimator was developed and applied during World War II by Turing and Good. The main focus of their work was in cracking the Enigma military code used by Germany and its allies. The Enigma was a type of enciphering machine used by the German armed forces to send messages securely.

In our notations Good-Turing estimator can be presented in following way:

$$\pi_k = \frac{1}{n}(k+1)\frac{n_{k+1}}{n_k}, \quad \pi_0 = \frac{n_1}{nn_0} \tag{17}$$

where $\pi_k$ the probability of a word occurring given that it appeared k times, $n_k$ denote the mathematical expectation of the number of item types that occur exactly k times. A useful part of Good-Turing methodology is the estimate that the total probability of all unseen objects is $p_0 = n_1/n = n_1/(n_1+2n_2+...+kn_k+...)$, from which follows(13). It is important to note that the probabilities $p_0$ and $\pi_0$ are not the same!

• It is easy to note, that estimator (15) (Mnatsakanian and Mirzoyan, 1988) in the particular case $r = 0$ (see (16)) coincides with the Efron-Thisted estimator derived in their famous article "Estimating the Number of Unseen Species: How Many Words Did Shakespeare Know?" (Efron and Thisted, 1976): end

$$S(\tau) = \sum_{k=1}^{\infty}(-1)^{k+1}(\tau/T)^k n_k \tag{18}$$

where $\tau = t-T$ is the prediction time interval/prediction size, $S(T) = n_0(T) - n_0(t)$ is the expected number of flare stars/new species that will be observed in a prediction interval of length/prediction sample of size $\tau$. Another useful prediction estimator is proposed by Solow and Polasky (1999), in which Ambartsumian estimator is figured:

$$S(m) = \frac{n_1^2}{2n_2}\left[1 - \left(1 - \frac{2n_2}{n_1 n}\right)^m\right], \tag{19}$$

where m and n are prediction size (number of new flares) and sample size (total number of flares), S(m) has same meaning as S(τ) - expected number of new flare stars. If m < $nn_1/2n_2$ (in practice, usually this is the case), then (19) can be simplified up to

$$S(m) \simeq \frac{mn_1}{n}. \tag{20}$$

- Earlier analogue of the estimator (14) was proposed by Harris (1968) in one of his study on classical occupancy problem.

## 5. Concluding Remarks

Considered astronomical estimators (mainly Ambartsumian estimator) repeatedly and successfully have been applied in statistical studies of ordinary flare stars (e.g., Mirzoyan(1981); Akopian and Parsamian (2009)), X-ray flare stars (e.g., Pye et al. (2015)), solar type stars with superflares (Akopian, 2013). It is likely that in the near future (with the accumulation of the necessary data), they can be successfully applied in studies of supernova stars and gamma bursts. Using by his own estimator Ambartsumian estimated the total number of stars the Pleiades cluster and came to important cosmogonical conclusion that the stage of flare activity is a regular stage in the evolution through which all red dwarf stars must pass.

Unfortunately works of astronomers in this field for a long time were unknown among the statisticians. In his letter to editor of journal "The Observatory" named "From Shakespeare to the Pleiades via statistics" Kiang (1987) wrote "…I was enchanted by the Odeto Statistics in the 1986 January 24 issue of "Science", and readers of these pages may like to know that a variation of the same theme was sung some years ago by the Soviet astrophysicist V. A. Ambartsumian in a heavenly guise. The common theme is the estimation of the number of no events from the numbers of single events, double events, triple events, and so on". The situation was not changed up to now. In return, the astronomers also were not familiar with the results of statisticians in this field. As a result, some identical results are obtained in both areas of science, sometimes with a significant time lag. Such a situation could be explainable and understandable until the end of the 20th century, but it is abnormal in the 21st.

**Appendix A**

We must to show that in heterogeneous (the mean frequencies of flares are varying indifferent flare stars) case estimator (13) gives the lower boundary. If the mean frequencies of flares are not the same then

$$p_k = \int \varphi(\nu) \frac{(\nu t)^k e^{-\nu t}}{k!} d\nu$$

$$n_k = N p_k$$

where φ(ν) is a density distributions of the mean frequencies of flares. Note that expression(13) one can write down as:

$$n_0(2n_2 + 3n_3 + ... + k n_k + ...) = n_1(n_1 + n_2 + ... + n_{k-1} + ...)$$

It is obvious that is enough to show that

$$k n_0 n_k \geq n_1 n_{k-1}$$

For this purpose we use Hölders inequality:

$$\left| \int f(\nu) g(\nu) d\nu \right|^k \leq \int |f(\nu)|^k d\nu \cdot \left[ \int |g(\nu)|^{\frac{k}{k-1}} d\nu \right]^{k-1}$$

As a first step we put as

$$f(\nu) \equiv \left[ e^{-\nu t} \varphi(\nu) \right]^{\frac{1}{k}}$$

$$g(\nu) \equiv \left[ e^{-\nu t} (\nu t)^k \varphi(\nu) \right]^{\frac{k-1}{k}}$$

From Hölder's inequality follows

$$\int \varphi(\nu) e^{-\nu t} d\nu \cdot \left[ \int \varphi(\nu) e^{-\nu t} (\nu t)^k d\nu \right]^{k-1} \geq \left[ \int \varphi(\nu) e^{-\nu t} (\nu t)^{k-1} d\nu \right]^k \quad (21)$$

As a second step we put as

$$f(\nu) \equiv \left[e^{-\nu t}(\nu t)^k \varphi(\nu)\right]^{\frac{1}{k}}$$

$$g(\nu) \equiv \left[e^{-\nu t}\varphi(\nu)\right]^{\frac{k-1}{k}}$$

From Hölder's inequality follows

$$\int \varphi(\nu)(\nu t)^k e^{-\nu t} d\nu \cdot \left[\int \varphi(\nu) e^{-\nu t} d\nu\right]^{k-1} \geq \left[\int \varphi(\nu) e^{-\nu t}(\nu t) d\nu\right]^k \quad (22)$$

Multiplying (21) and (22) one can obtain

$$\int \varphi(\nu)(\nu t)^k e^{-\nu t} d\nu \cdot \int \varphi(\nu) e^{-\nu t} d\nu \geq \int \varphi(\nu) e^{-\nu t}(\nu t) d\nu \cdot \int \varphi(\nu) e^{-\nu t}(\nu t)^{k-1} d\nu$$

or

$$k p_0 p_k \geq p_1 p_{k-1}$$

$$k n_0 n_k \geq n_1 n_{k-1}$$

The proof completed.